# Constraints on the number of X-ray Pulsars in IC 10 from a deep *XMM-Newton* Observation

Jun Yang*[1] | Silas G. T. Laycock[2] | Daniel R. Wik[1]

[1]Department of Physics and Astronomy, the University of Utah, Utah, USA

[2]Department of Physics and Applied Physics, University of Massachusetts, Lowell, Massachusetts, USA

**Correspondence**
*Jun Yang. Email: junyang@astro.utah.edu
www.jyang.us

**Present Address**
115 S 1400 E, Rm 201, Salt Lake City, Utah, 84112.

We report the most sensitive search yet for X-ray pulsars in the dwarf starburst galaxy IC 10, which is known to contain a population of young high mass X-ray binaries. We searched for pulsations in 207 point-like X-ray sources in the direction of IC 10 by a 2012 *XMM-Newton* observation with a total exposure time of 134.5 ks. Pulsation searches in faint objects can be sensitive to the energy bands of the light curves, and the source and background extraction areas. We analyzed separately the PN and MOS barycenter corrected 0.2-12 keV data, with good time interval filtering. Different schemes for source and background extraction were compared, and the search was repeated in the narrower 0.5-8 keV energy band to increase the signal-to-noise ratio. For the most conservative parameters, 5 point sources produced significant peaks in the Lomb-Scargle periodogram (99% significance, single trial, assuming white noise). A similar number of different candidates result from alternative analyses. A ~4100 s period seen in all 3 instruments for the black hole (BH) + Wolf-Rayet (WR) binary IC 10 X-1 is probably due to red noise of astrophysical origin. Considering the periods, luminosities, and spatial distribution of the pulsar candidates in the direction of IC 10, they do not belong to the same distribution as the ones in the Magellanic Clouds and Milky Way. This result holds even if the candidates are spurious, since if the Small Magellanic Cloud were placed at the distance of IC 10, we would expect to see ~5 pulsars at $L_x > 10^{36}$ erg s$^{-1}$ inside the $D_{25}$ contour, and their periods would be of order 100 seconds, rather than the mostly ~1 s periods for the candidates reported here, which lie outside the main body of the galaxy.

## 1 | INTRODUCTION

X-ray binaries (XRBs) are a powerful tool for studying the formation and evolution of neutron stars (NSs) and black holes (BHs). Constraining their formation rate and evolutionary channels is key for understanding the X-ray emission from high-z galaxies, their importance for feedback in the early Universe (e.g. Fragos et al., 2013a), and the populations of gravitational wave progenitors (e.g. Abbott et al., 2016). Population synthesis models predict that the formation efficiency of XRBs strongly depends on stellar age and metallicity (e.g. Fragos et al., 2013b; Linden et al., 2010), but remain sparsely tested owing to the difficulties in understanding both of these quantities for Galactic systems and the precise ages of systems that can be observed in the Magellanic clouds.

The Local Group dwarf irregular galaxy IC 10 provides a unique opportunity to study compact object formation at early ages. It is the nearest starburst galaxy to us located at a distance of 660 ± 60 kpc, and is undergoing vigorous star formation activity. It hosts an unusually high number of Wolf-Rayet (WR) stars (Crowther et al., 2003) and XRBs, but, crucially, a dearth of red supergiants limits the age of the starburst to 10 Myrs (Massey et al., 2007; Wilcots & Miller, 1998). This, in turn, limits the XRB compact object progenitors to about $M \geq 12 M_\odot$. With a metallicity of about 0.15 solar, comparable to the Small Magellanic Cloud (SMC) (Crowther et al., 2003), IC 10 is an ideal laboratory for studying XRBs and their optical counterparts for a well-constrained progenitor sample, with the ultimate goal of providing a critical benchmark and test of models of compact object formation and evolution.

Currently, the galaxy produces stars at a rate of 0.5 $M_\odot$/yr (Leroy et al., 2006; Massey & Holmes, 2007), which implies that the gas supply can last for just a few more billion years. This is a much higher star formation rate per unit molecular ($H_2$) or total gas content ($H + H_2$) than in most other galaxies (Leroy et al., 2006). For these reasons, IC 10 is the best example of a low-mass, metal-poor, starburst galaxy in the Local Group, acting as a young analog of the Small Magellanic Cloud.

The X-ray binary population of IC 10 was inferred by Wang et al. (2005) based on the spatial clustering of point sources within the optical outline of the galaxy. Furthermore, the integrated spectrum of these objects follows an "High Mass X-ray Binary (HMXB)-like" hard power-law. Monitoring over several years with *Chandra* has revealed a sample of 110 unique point sources, 21 of which are variable (Laycock et al., 2017a) and up to 16 are associated with blue supergiants (with an overlap of 5 objects between those subgroups) (Laycock et al., 2017b). Perhaps the single strongest candidate for a pulsar hosting system is IC 10 X-2, a large amplitude X-ray transient paired with an extremely luminous blue supergiant (Kwan et al., 2018; Laycock et al., 2014). Unfortunately this object was not in outburst during the *XMM-Newton* observation. The most prominent source, IC 10 X-1, has been intently scrutinized (e.g. Barnard et al., 2008; Prestwich et al., 2007; Steiner et al., 2016) and is believed to be a black hole + WR binary. Pasham et al. (2013) reported a mHz quasi-periodic oscillation (QPO) in X-1 in the same dataset analyzed here.

Noori et al. (2014, 2017) conducted a deep radio pulsar search of IC 10, and were unable to identify any continuous pulsed signals. From the null result of their search, they claim that there are few (if any) normal radio pulsars in IC 10, a curious result given the large number of bright young core collapse supernova remnants.

The pulsar candidates in the much younger galaxy IC 10 can form a comparison sample for the Magellanic Cloud X-ray pulsars. The SMC is a dwarf irregular galaxy at a distance of 62 kpc from the Milky Way (Graczyk et al., 2014; Scowcroft et al., 2016). It contains a large and active population of HMXBs (e.g. Christodoulou et al., 2016.1; Coe & Kirk, 2015; Haberl & Sturm, 2016; Klus et al., 2014; Galache et al., 2008; Townsend et al., 2011; Yang et al., 2017a,1).

In this paper, we present a deeper survey of the X-ray sources in the direction of IC 10 using data from the *XMM-Newton* Science Archive (XSA). We have conducted a complete study of the properties of these sources, including their light curves and spectra, as well as searched for pulsations. We have combined the 2012 data with a catalog of IC 10 sources from the *Chandra* X-ray archive (Laycock et al., 2017a), and the previous *XMM-Newton* observation of the galaxy (Wang et al., 2005). The results from this study will help us better understand the accretion process in compact binaries, the mass transfer from WR stars to their companions, as well as tidal interactions and gravitational wave emission from a young starburst environment, which is expected to be heavily populated by NS and BH binaries born in roughly the last 10 million years. The star formation rate in this active environment can be compared with that in the Milky Way, the Magellanic Clouds, and other nearby galaxies in order to produce a map of stellar populations of varying ages in the galaxies of the Local Group.

This paper is organized as follows: in § 2 and 3, we briefly discuss the deep 2012 *XMM-Newton* observation of IC 10 and our analysis method, and in § 4, we describe the results of our analysis. Finally, we discuss our results in § 5 and conclude in § 6.

## 2 | OBSERVATIONS

Our IC 10 dataset was obtained by *XMM-Newton* in August 18-20, 2012, during a single observation with a total exposure time of 134.5 ks performed with the European Photon Imaging Camera (EPIC). Both PN and Metal Oxide Semi-conductor (MOS) instruments were active, and were used in this study.

All of the point sources in the 2012 *XMM-Newton* field of view are shown in Figure 1 . The background is the RGB PN image combined from red 0.2-1 keV, green 1-3 keV and blue 30-8 keV. White circle shows the 4′ D25 radius of IC 10. The sources numbered in green circles are those with candidate pulsations. The yellow and cyan circles are the background areas used for the PN light curve extractions.

There are 207 X-ray point sources in the XSA point-source catalog for this dataset.

The distribution of these source photon counts in the histogram is an order of magnitude above the logN-logS curve for the background AGN population (e.g. Kim et al 2004). There are 29 sources inside of the D25 radius of IC 10. The photon counts of these 29 sources are from ∼25 to ∼28000.

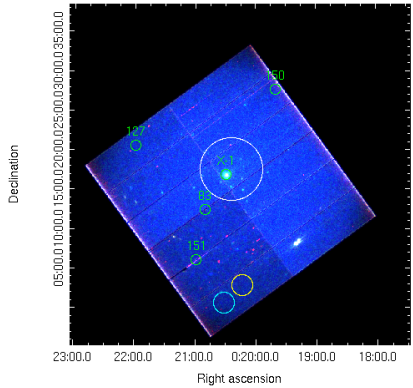

**FIGURE 1** IC 10 *XMM-Newton* image shows the 207 detected sources. The sources with numbers labeled in green are the pulsar candidates detected by PN while use 40″ radius as the source extraction size. The white circle shows the galaxy IC 10 outline with the D25 radius of 4′ (Huchra et al., 1999), which is centered at RA 00h:20m:23.16s and Dec +59°:17′:34.7″. The yellow and cyan circles are the 1st and 2nd background regions, respectively, used to subtract the PN light curves.

## 3 | ANALYSIS

We analyzed all of the 207 X-ray point-sources in the direction of IC 10 using the latest calibration files provided by the Science Analysis System (SAS)[1] with the High Energy Astrophysics Science Archive Research Center Software. Based on the source list from the XSA pipeline, we have generated source event files, high time-resolution (0.0734 s) light curves, periodograms, and spectra.

Since some of the X-ray sources may show short periodic pulsations, we applied a barycentric correction (BC) to all the data using the SAS command *barycen*. The arrival time of each photon was shifted to the barycenter of the Earth-Sun system from the position of the orbiting telescope. The PN light curve and the spectrum of each source were extracted with EPIC event pattern from 0 to 4 (singles and doubles) and with the PN flags XMMEA_EP. The MOS data were filtered with pattern from 0 to 12 (taking good events, singles, doubles, triples and quadruples) and with the standard MOS flags XMMEA_EM. In the reductions, we excluded the energy band 12-15 keV because of high background flares.

The good time interval (GTI) file is produced by *tabgtigen* with count rate $\leqslant 0.4$ to the PN light curves. The time series of the various light curves, with GTI filter applied, are not continuous, so we used the Lomb-Scargle (LS) method (Scargle, 1982; Yang et al., 2017b) to analyze their periodograms.

[1] http://xmm.esac.esa.int/sas/

The algorithm modifies the standard periodogram formula and finds a time delay such that the sinusoids are mutually orthogonal at sample times and are adjusted for the potentially unequal powers of the two basis functions. This results in a better estimate of the power at each chosen frequency (Press et al., 1992). The LS method is equivalent to Lomb's least-squares method (Lomb, 1976). The significance (sig) is heretofore calculated from the number $M$ of independent frequencies and the LS power $P_X$: $[1 - M \exp(-P_X)]100\%$ (Press et al., 1992; Yang et al., 2017b).

Several extraction schemes were employed in order to explore the effects of the choice of energy-range, and extraction regions for source and background. These produced differing results as discussed in the next section.

We extracted light curves in two energy bands: broad (0.3-12 keV) and narrow (0.5-8 keV), the latter designed to increase signal-to-noise ratio, chosen after examining the XSA pipeline products, which show that most of the photon counts are located in the energy band 0.5-8 keV.

Initially, annulus background subtraction (ABS) was performed using the counts within a concentric annular region with inner and outer radii of 50″ and 100″, respectively. This method proved unworkable as source contamination occurred in the background regions of many sources. Next, following the approach of Israel et al. (2016), raw light curves were extracted from a circular region of 40″ radius centered on each point source, with background light curves extracted in a separate clean region (See Figure 1). This radius of extraction closely approximates the 90% encircled energy contour across the EPIC field, and ensures compatibility with the results of the Israel et al. (2016) serendipitous pulsar survey.

We also extracted light curves using a smaller 20″ region, to reduce source confusion. Again producing light curves using both annular background regions, and fixed background regions.

The choice of source and background region has a noticeable influence in the determinations of the quantities associated with the light curves. When we subtract a fixed background for all of the sources in IC 10, the signal-to-noise ratios are different from those obtained from the annulus background subtraction, and the pulsations of different sources appear with sigs > 99%.

## 4 | RESULTS

### 4.1 | Light Curves and Periodograms

Using a fixed background subtraction (FBS), as shown in the yellow region in Figure 1 , we find X-1, source 127 and source 151 with significant pulsations in the broad energy band of 0.2-12 keV (Table 1 ). Source 127 with possible spin period of 0.84 s is very faint with 949 counts in the 0.2-12 keV energy band. It is located at the far outskirt of IC 10 and more likely is a flare star. Source 151 with 3824 counts was found to have pulsations of 5233 s, which might be instrumental in origin, as it is located near the CCD detector gap. The peak frequency might also be due to the red noise as it is near the edge of the low frequencies. Moreover, the RGB image Figure 1 provides immediately some spectral information via the magenta color for this source and shows it as soft source.

In the narrow energy band (0.5-8 keV), we found that source 83 and source 150 have pulsations of 1.25 s and 1.04 s, respectively. Source 83 is near the CCD gap with few counts of 125. Source 150 is very faint with 378 counts and far at the edge of the PN detector, so the pulsations might be affected by the instruments.

We find that when we chose a different fixed background region (the cyan circle in Figure 1) to subtract from all the light curves, the pulsations of source 150 disappear, which suggests the periodicities might be chip related. However, X-1 and Source 83 still show pulsations with sigs > 99%, as shown in Table 1. Source 83, with a much higher counts of 2504 comparing to the first background region we chose, still shows the same period of 1.25 s. Again, since this source is located near the edge of the CCD gap, its periodicity might be suspicious.

According to Figure 5 in Yang et al. (2017b), the minimum EPIC PN photon counts to detect the pulsations are ∼ 70. We use the same pulsation search method and script as that in Yang et al. (2017b), and, therefore, can claim that the pulsations from these sources (with photon counts > 100) are promising.

### 4.2 | X-1 from MOS

We also analyzed the time series extracted from MOS detections using the 0.5-8 keV energy band with FBS and GTI filter (Table 1). In the MOS1 dataset, we did not find high-sig pulsations except in the case of X-1. Its pulse period is 4136 s. In the MOS2 dataset, we found pulsations for X-1 with the same pulse period of 4136 s.

We did not find pulsations in source 127, 83 and 150 with the MOS data since their time resolution is lower than the periodicities found from the PN detections. Source 151 does not show pulsations in MOS, which confirms that the period found from PN might be due to the instrumental effect or red noise.

The pulsation period of X-1 lies in the part of the periodogram that is dominated by red noise and it will need an independent verification. The average spin period for the values listed in Table 1 is 5066 s, with a spread of ±21%. All three cameras have observed a similar periodicity, which also appear in various subsets of the original dataset. This strengthens the view that we have not observed a random fluctuation in the noise. The term "noise" can refer to intrinsically disordered variability of the source (stochastic variability). If three cameras all observe at the same time, they would detect the same noise features. A genuine periodic signal must be seen in independent observations on different dates. If the compact object of X-1 turns out to be a NS, that would be consistent with the conclusion of Laycock et al. (2015) who consider a low-mass companion to the WR star (although they could not rule out a low-mass BH); but it would be at odds with the conclusion of Steiner et al. (2016), who argue in favor of a BH based on the observed X-ray spectrum of the source.

## 5 | DISCUSSION

All of the sources identified as candidate pulsators (except X-1) with extraction radius of 40″ are projected at the far outskirts of IC 10 and their actual locations are therefore most likely Galactic. No candidate lies within the D25 contour. This remains true no matter which of our extraction methods is adopted. This is an interesting result, and in accordance with the dearth of radio pulsars noted by Noori et al. (2017).

According to the distribution of counts for positive detection in searches for pulsations in *XMM-Newton* data (Yang et al., 2017a), a minimum of ∼70 counts are needed. The sources 83, 127, 150, and 151 were detected with more than 70 photons counts. However, we have searched 207 sources in total. Therefore with the > 99% sig, the statistical expectation is for pulsations to be found in ∼2 sources with false detections, per extraction scheme. Since we searched two energy bands, with two different background regions, we can expect 0.01 × $207 \times 2 \times 2 = 8$ false positives. The results for our alternative analyses using smaller extraction regions, and/or annular backgrounds produced a similar number of (different) candidates. This implies our pulsar candidates are consistent with the actual false positive rate.

**TABLE 1** EPIC PN Sources (0.2-12 keV)

| Source | Net Counts | Period (s) | sig (%) |
|---|---|---|---|
| EPIC PN Sources (0.2-12 keV) | | | |
| X-1 | 46899 | 4215± 5 | 100 |
| 127 | 949 | 0.84±0.00 | 99.35 |
| 151 | 3824 | 5233±10 | 99.92 |
| EPIC PN Sources (0.5-8 keV) | | | |
| X-1 | 43528 | 6423 ±12 | 100 |
| 83 | 125 | 1.25 ±0.00 | 99.33 |
| 150 | 378 | 1.04 ±0.00 | 99.64 |
| with alternate background region | | | |
| X-1 | 45907 | 6422 ±12 | 100 |
| 83 | 2504 | 1.25 ±0.00 | 99.15 |
| MOS Sources with FBS (0.5-8 keV) | | | |
| X-1 from MOS1 | 15874 | 4136±5 | 99.99 |
| X-1 from MOS2 | 16067 | 4136±5 | 100 |

That IC 10 hosts a significant HMXB population seems probable based on the work of Wang et al. (2005), and Laycock et al. (2017a,1), therefore the lack of detected pulsars implies a true absence, perhaps because they have not had time to form. Whether or not our dataset has the sensitivity to make this claim can be examined as follows.

The SMC pulsar population covers the luminosity range of $L = 10^{31.2}$–$10^{38}$ erg s$^{-1}$. If the population of the IC 10 pulsars was the same as that of the SMC, we estimate how many should have been detected by placing the pulsars in the SMC at the distance ($D$) of IC 10. If we set the pulsar luminosity ($L$) equal to $10^{37}$ erg s$^{-1}$, the flux ($F$) of the pulsar is

$$F = \frac{L}{4\pi D^2} = \frac{10^{37} \text{ erg s}^{-1}}{4\pi(660 \text{ kpc} \times 3 \times 10^{21} \text{ cm kpc}^{-1})^2} \quad (1)$$
$$\approx 2.0 \times 10^{-13} \text{ erg s}^{-1} \text{ cm}^{-2}.$$

Then we used the tool WebPIMMS[2] (Portable, Interactive, Multi-Mission Simulator) to convert the count rate from the *XMM-Newton* PN flux calculated from Equation (1). The input energy range was 0.3–8 keV and the output energy range was 0.3–8 keV. We used a power-law model with Galactic nH $5.03 \times 10^{21}$ cm$^{-2}$ (calculated using NASA's HEASARC tool[3] for the SMC sightline), Redshift was set as none, and Intrinsic nH as 0.0 cm$^{-2}$. The predicted *XMM-Newton* PN THIN count rate is about $4.2 \times 10^{-2}$ counts/s. Thus, with an exposure time of ~100 ks, the exposure time of this deep *XMM-Newton* observation for IC 10 within the GTI, the expected photon counts from the pulsar should be $4.2 \times 10^{-2}$ counts/s $\times$ 100 ks = 4200 counts. Similarly, if the SMC pulsar with $L = 10^{36}$ erg s$^{-1}$ or $10^{35}$ erg s$^{-1}$ is in outburst and located at the distance of IC 10, the expected photon counts for this pulsar would be 420 or 42, respectively. Therefore, we expect to detect pulsations for typical type I outburst.

The expected number of pulsars to be detected depends on the population and the duty cycle. Here the meaning of duty cycle is the fraction of time the pulsar spent above the detectable luminosity cutoff $L = 10^{35}$ erg s$^{-1}$. Currently ~60 pulsars are found in the SMC (Haberl & Sturm, 2016; Yang et al, 2017b). If we set their duty cycle as 10% (Wang & Gotthelf, 1998), $6 \pm 3$ pulsars would be expected in outburst, assuming root-N Poisson statistics. This can be compared to zero actual detections.

A possible explanation of its far fewer confirmed Be/X-ray binaries (Be-XBs) is the relation between the number density of Be-XBs and the recent star formation activity in each galaxy. Antoniou et al. (2010) found that the ages of the stellar populations in which the Be-HMXBs are embedded peak at ~25-60 Myr in the SMC compared to a younger population in the LMC (~6-25 Myr old; Antoniou & Zezas, 2016). This could explain why currently there are fewer pulsars (only 16) found in the Large Magellanic Cloud (LMC). The population in IC 10 is even younger than LMC, so this trend would be consistent with fewer pulsars.

## 6 | CONCLUSION

During the full search of the deep *XMM-Newton* 2012 observation in the direction of IC 10, 207 sources from PN and MOS were analyzed with different energy bands and with various extraction and background subtraction schemes. For the most conservative extraction scheme (following Israel et al., 2016), 5 candidates are found, which is consistent with the false positive rate.

Sources identified as pulsator candidates include source 151 as well as X-1 (likely red noise related). Among the fainter objects, 3 candidates (sources 83, 127 and 150) with pulsations were found with > 99% sig. With FBS, sources 127 and 151 were detected pulsations of 0.84 s and 5233 s, respectively, within the energy band of 0.2-12 keV. Aiming to increase the signal-to-noise ratio, we searched for the pulsations within the 0.5-8 keV energy band. Source 83 and 150 show periodicities. Using a different background region, the pulsations from source 150 disappear.

Due to their locations however, from the RGB image shown in Figure 1 , sources 83, 127, 150, and 151 are more likely foreground (Galactic) stars. The XSA data show they have soft spectra, also typical of stellar coronal sources. Whether or not the pulsations from any of these sources turn out to be real, the pulse period distribution of these candidates (which are mostly very short) are different from the X-ray pulsars in the SMC and LMC, which cluster in the 50-200 second range. The short period Magellanic systems only appear as rare transient outbursts at very high luminosity (e.g. SMC X-2, Kennea et al. 2016; SMC X-3, Tsygankov et al. 2017, Weng et al. 2017, Townsend et al. 2017; A0538, Skinner et al. 1982). Hence, regular monitoring observations of IC 10 are strongly needed to enable targeted followup to search for such fast pulsars. The growing number of Supergiant Fast X-ray Transients (e.g. Vasilopoulos et al. 2018) and ultra-luminous X-ray pulsars (e.g. Carpano et al. 2018) found in star-forming regions also motivates this course of action.

---

[2] https://heasarc.gsfc.nasa.gov/cgi-bin/Tools/w3pimms/w3pimms.pl
[3] https://heasarc.gsfc.nasa.gov/cgi-bin/Tools/w3nh/w3nh.pl

With high photon counts, the X-ray binary IC 10 X-1 was found to have a pulsation period of ∼1.4 hr from PN, MOS1, and MOS2. Pasham et al. (2013) found the power spectrum of X-1 is dominated by a steep red-noise distribution, with a mHz QPO superimposed. Thus the existence of serial correlations in the light-curve at longer timescales is far from unexpected. On the contrary this feature could be related to intrinsic flaring from the accretion disk or hotspot.

Therefore although the young dwarf starburst galaxy IC 10 seems to host a population of HMXBs in agreement with the Grimm et al. (2003) scaling law; the species of HMXBs differ from the Magellanic and Galactic populations in the following ways: (1) Supergiant HMXBs dominate instead of Be-HMXBs, (2) pulsars are either absent, or have a starkly different distribution of periods. Moreover, the age of the galaxies play a significant role in the formation of HMXBs. IC 10 has a much younger age and a comparable metallicity to the SMC, so potentially it has so far produced fewer pulsars.

# REFERENCES

Abbott, B.P., Abbott, R., Abbott, T.D., Abernathy, M.R., Acernese, F., Ackley, K., Adams, C., Adams, T., Addesso, P., Adhikari, R.X. and Adya, V.B., 2016. The Astrophysical Journal Letters, 818(2), p.L22.

Antoniou, V., & Zezas, A. (2016). MNRAS, 459, 528.

Antoniou, V., Zezas, A., Hatzidimitriou, D., & Kalogera, V., 2010, ApJL, 716, L140

Barnard, R., Clark, J. S., & Kolb, U. C. 2008, A&A, 488, 697

Carpano, S., Haberl, F., Maitra, C., & Vasilopoulos, G. 2018, MNRAS, 476, L45

Christodoulou, D. M., Laycock, S. G. T., Yang, J., & Fingerman, S. 2016, ApJ, 829, 30

Christodoulou, D. M., Laycock, S. G. T., Yang, J., & Fingerman, S. 2017, Research in Astronomy and Astrophysics, 17, 059

Coe, M. J., & Kirk, J., 2015, MNRAS, 452, 969

Crowther, P. A., Drissen, L., Abbott, J. B., Royer, P., & Smartt, S. J. 2003, A&A, 404, 483

Fragos et al. 2013a, ApJ, 764, 41

Fragos et al. 2013b, ApJ, 776L, 31

Graczyk, D., Pietrzyński, G., Thompson, I. B., et al 2014, ApJ, 780, 59

Grimm, H.-J., Gilfanov, M., & Sunyaev, R. 2003, MNRAS, 339, 793

Haberl, F., & Sturm, R. 2016, A&A, 586, A81

Huchra J. P., Vogeley, M. S., and Geller, M. J., ApJS, 121, 287H

Israel, G. L., Esposito, P., Rodríguez Castillo, G. A., Sidoli, L., 2016, MNRAS, 462, 4371

Kennea, J., Coe, M. J.; Laycock, S. G. T., et al. 2016, HEAD, 1512012K

Kim, D.-W., Wilkes, B. J.; Green, P. J., et al. 2004, ApJ, 600, 59K

Klus, H., Ho, W. C. G., Coe, M. J., Corbet, R. H. D, & Townsend L. J., 2013, MNRAS, 437, 3863

Galache J. L., Corbet R. H. D., Coe M. J., Laycock S., Schurch M. P. E., Markwardt C., Marshall F. E., Lochner J., 2008, ApJS, 177, 189

Kwan, S., Lau, R. M., Jencson, J., et al. 2018, ApJ, 856, 38

Laycock, S., Cappallo, R., Williams, B. F., et al. 2017, ApJ, 836, 50

Laycock, S., Christodoulou, D. M., Williams, B. F., Binder, B., and Prestwich, A. 2017, ApJ, 836, 51

Laycock, S. G. T., Maccarone, T. J., & Christodoulou, D. M. 2015b, MNRAS, 452, L31

Laycock, S., Cappallo, R., Oram, K., Balchunas, A. 2014 ApJ, 789, 64

Leroy A., Bolatto, A., Walter, F., & Blitz, L. 2006, AJ, 643, 825

Linden et al. 2010, ApJ, 725, 1984

Lomb, N. R. 1976, Ap&SS, 39, 447

Massey, P., Olsen, K. A. G., Hodge, P. W., Jacoby, G. H., McNeill, R. T., Smith, R. C., Strong, S. B. 2007, AJ, 133, 2393

Massey, P., and Holmes, S., 2002, ApJ, 580, L35

Morrison, R., & McCammon, D. 1983, ApJ, 270, 119

Noori, H. A., Roberts, M., Champion, D., et al. 2017, arXiv:1702.08214

Noori, H. A., Roberts, M., Champion, D., et al. 2014, AAS, 22315315N

Pasham, D. R., Strohmayer, T. E., & Mushotzky, R. F. 2013, ApJ, 771, 44

Press, W. H., Teukolsky, S. A., Vetterling, W. T., & Flannery, B. P. 1992, Numerical Recipes in C, Cambridge: Cambridge Univ. Press

Prestwich, A. H., Kilgard, R., Crowther, P. A., et al. 2007, ApJ, 669, 21

Scargle, J. D. 1982, AJ, 263, 835

Scowcroft, V., Freedman, W. L., Madore, B. F., et al. 2016, ApJ, 816, 49

Steiner, J. F., Walton, D. J., García, J. A. 2016, ApJ, 817, 154

Skinner, G. K., Bedford, D. K., Elsner, R. F. et al. 1982, Nature, 297, 568-570

Townsend, L. J., Coe, M. J., Corbet, R. H.D., & Hill, A. B., 2011 MNRAS, 416, 1556

Townsend, L. J., Kennea, J. A., Coe, M. J., et al. 2017, MNRAS, arXiv170102336T

Tsygankov, S. S., Doroshenko, V., Lutovinov, A. A. et al. 2017, A&A, arXiv170200966T

Vasilopoulos, G., Maitra, C., Haberl, F., Hatzidimitriou, D., & Petropoulou, M. 2018, MNRAS, 475, 220

Wang, Q. D., & Gotthelf, E. V. 1998, ApJ, 509, L109

Wang, Q. D., Whitaker, K. E., & William, R. 2005, MNRAS, 362, 1065

Weng, S.; Ge, M.; Zhao, H.;. 2017, ApJ, arXiv170102983W

Wilcots, E. M., & Miller, W. 1998, AJ, 116, 2363

Yang, J., Laycock, S. G. T., Drake, J. J., et al. 2017, Astron. Nachr./AN, 338, Issue 2-3, 220-226

Yang, J., Laycock, S. G. T., Christodoulou, D. M., et al. 2017, ApJS, 839, Number 2,119

Yang, J., Zezas, A., Coe, M. J., et al. 2018, MNRAS: Letters, 479, Issue 1, L1-L6